\documentclass[aps,prl,showpacs,twocolumn]{revtex4}

\usepackage{graphicx}% Include figure files
\usepackage{dcolumn}% Align table columns on decimal point
\usepackage{bm}% bold math

\begin{document}

%\preprint{APS/123-QED}

%\preprint{}

%\title{Efficient free energy profile reconstruction \\using adaptive stochastic perturbation protocols}
\title{Efficient free energy profile reconstruction\\using adaptive stochastic perturbation protocols}

\author{Ognjen Peri\v si\'c and Hui Lu}
\email[]{huilu@uic.edu}
%\thanks{}
\affiliation{Department of Bioengineering, University of Illinois at Chicago, Chicago, IL 60607}

\date{\today}
\begin{abstract}
Application of Jarzynski nonequilibrium work relation to free energy calculation is limited by the very slow convergence of the estimate when dissipation is high. We present a novel perturbation protocol able to improve the convergence of Jarzynski estimator when it is applied in the reconstruction of the potential of mean force. The improvement is based on the application of the adaptive external work variation in addition to the one caused by thermal fluctuations.
\end{abstract}

\pacs{05.70.Ln, 82.40.Bj, 83.10.Mj, 87.64.Aa, 87.15.La }
%\keywords{}

\maketitle

%\section{}
%\subsection{}
%\subsubsection{}
Mechanical properties of bio-polymers, e.g. proteins and nucleic acids, often determine their functioning and play a significant role in the interactions they have with other biomolecules. Single molecule manipulation experiments using atomic force microscopy (AFM) \cite{sppclll,rgofg97}, optical tweezers \cite{ldstc2002} and $\it in-silico$ methods such as steered molecular dynamics (SMD) \cite{hl1998,ls1999,sppclll} opened a possibility for the analysis of those properties. The mechanical resistance measured during a single molecule stretching experiment is determined by the molecule's free energy profile, also called potential of mean force (PMF). The knowledge of this profile is therefore essential for the understanding of the biopolymers' mechanical behavior.

The second law of thermodynamics states that the average external work $\left\langle W \right\rangle$ used to perturb a given system between two states is always greater than or equal to the corresponding free energy difference, namely $\left\langle W \right\rangle\geq$ $\Delta F$. The equality applies only when the external perturbation is reversible \cite{fr1965}. The direct calculation of free energy difference is difficult in the non-reversible cases where an average dissipation is significant and unknown. That difficulty is additionally pronounced when PMF has to be calculated because the behavior of the dissipation changes along the reaction coordinate \cite{ps2004}.

In 1997 C. Jarzynski presented a theoretical framework able to cope with the problem of free energy calculation in the form of the nonequilibrium work relation \cite{cj1997,gc1999}. This relation gives a direct connection between the exponential average of the external work used to move a system between two equilibrium states and the exponential value of the corresponding free energy difference
\begin{eqnarray}
\left\langle e^{-\beta W}\right\rangle = e^{-\beta \Delta G}.
\end{eqnarray}
The most important property of this equality is that the work performed on a given system does not have to be reversible. It is satisfied for any perturbation, close to equilibrium or far from it \cite{cj1997,grb2003}.

When an external perturbation is far from equilibrium and the number of work samples is limited, the Jarzynski based PMF estimate contains a bias \cite{grb2003,ps2004}. The bias can be seen in the estimate based on the numerical Brownian simulation of a single molecule constant velocity stretching experiment (Fig. 1) \cite{hs2001}. The average work based on 2000 trajectories (Fig. 1, curve 2) in this experiment is significantly bigger than the underlying PMF (Fig. 1, curve 1) when pulling velocity is high. Only a modest improvement can be achieved using Jarzynski estimator with the same 2000 trajectories (Fig.~1,~curve~3). ~The reason behind the Jarzynski bias lies in the fact that nonequilibrium work relation emphasizes rarely occurring work samples with a small or negative dissipation~\cite{grb2003}.

The problem of the slow convergence of Jarzynski estimator \cite{grb2003} in the case of the normal single molecule constant velocity pulling experiments (normal pulling) can be overcome by the modification of the perturbation protocol. That modification should be able to increase the probability of generating work samples with a small dissipation. The reduction of the pulling velocity is an obvious way to achieve that, but it is often experimentally difficult or very costly \cite{ps2004}. The broader distribution of the external work can produce the same effect. This broadening can happen either unintentionally, e.g. due to the imperfections of the experimental setup, or it can be intentionally introduced through the additional random variation of the external work. The intentionally introduced work variation can be applied via symmetrically distributed random perturbation of the pulling cantilever/spring, i.e. through the external $\it noise$ with the mean value equal to zero. With this kind of the external perturbation, the measured work has larger variation but the same mean value as the work in the normal pulling experiments. Fig. 1 shows two Jarzynski PMF estimate (curves 4 \& 5) based on the constant velocity pulling with the additional external noise. Although calculated with the significantly smaller number of samples (both of those estimates are averages of 10 reconstructions, each based on 20 trajectories) those two estimates have much smaller bias due to increased work variation than the estimate coming from the normal pulling (Fig. 1, curve 3). In both cases, i.e. in normal pulling and in pulling with the additional $\it noise$, the exponential work averages along the pulling coordinate were calculated using weighted histogram protocol \cite{hs2001}. The external work was calculated using a constant velocity pulling assumption which is, obviously, only an approximation.
\begin{figure}[h]
\includegraphics{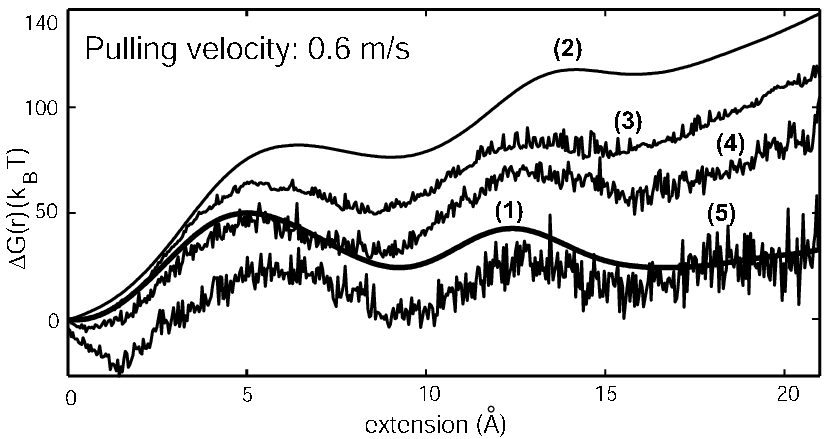}
\caption{\label{fig:figure_1}Original free energy profile compared to the reconstructions based on the normal pulling and stochastic pulling. (1) Original PMF; (2) $\left\langle W \right\rangle$; (3) Estimate based on normal pulling (4) Estimate, const velocity + noise, $\it m$=50; (5) Estimate, const. velocity + noise, $\it m$=80.}
\end{figure}

The thorough analysis of the influence of the additional noise on the behavior of Jarzynski-based PMF estimates in the single molecule manipulation experiments can be performed using Brownian motion formalism \cite{fr1965}. The trajectory of the pulled point is described using a discretized, one dimensional variant of Langevin equation $
r(r)=r(t-1)+\beta Df(r(t-1))\Delta t+\sigma _r \cdot \eta (t)$ which connects the position of a pulled point (which corresponds to the extension of a molecule in AFM and SMD experiments) to the force $\it{ f(r)}$ acting on it and to the random perturbation $\it\sigma _r \cdot \eta (t)$caused by the thermal fluctuations of the molecule and its environment \cite{fr1965}. The force $\it{ f(r)}$ is the first spatial derivative of the time-dependent Hamiltonian $ \it{ H(r,t)}=\Delta G(r)+\it{k(x(t)-r)}^2/2$, which describes a system made of a protein with the extension-dependent free energy profile $\Delta G(r)$ (Gibbs' free energy profile, Fig. 1, curve 1) and harmonic potential with the spring coefficient $\it{ k}$. The influence of the thermal environment is represented via normally distributed random perturbation $\sigma _r \cdot \eta (t)$ with a zero-mean and the variance $\left\langle \sigma _r ^2\right\rangle = 2 \it {D \Delta t}$. The quantity $\it D$ is the diffusion coefficient \cite{fr1965} and $\Delta t$ is the time step. The molecule is extended by the movement of the pulling point with the position $\it x(t)$ along the reaction coordinate. In this framework, the additional external noise can be introduced via instantaneous stochastic perturbation $\sigma _x \cdot \eta (t)$ of the pulling point: $\it x(t)=v \cdot t+\sigma _x \cdot \eta (t)$. This approach allows calculation of the external work using a constant pulling velocity approximation. For simplicity, we used a normal distribution to guide the external noise. An additional reason for the usage of this distribution is that according to the central limit theorem, the cumulative effect of any random signal follows Gaussian distribution.

The work generated by the random movement of the pulling point can be treated as a stochastic variable and expressed as a function of the instantaneous random variation of the spring extension $\it X$ and spring constant $\it k$, as $\it W=k\cdot X^2$. If $\it f_{X}(x)$ is the distribution function of the stochastic variable $\it X$, then the distribution function of the random work $\it W_R$ is ${f_{W{_R}}(w)}=\it (f_X(\sqrt{w/k})+f_X(-\sqrt{w/k}))/(2\cdot \sqrt{w\cdot k})$\cite{pa1991}. When the external noise is applied at the every time step of the simulation, its deviation can be expressed as a multiple of the random deviation of the pulled point using the multiplication factor $\it m$, $\it\sigma _x =m \cdot \sigma _r = m \cdot \sqrt{2 D \Delta t}$. The diffusion coefficient $\it D$ can be calculated from the slope of the average external work (e.g. Eq. 91-92 from \cite{ps2004}) but it is not required if one does not want to express $\it \sigma _x$ through $\it \sigma _r$. When a random movement of the pulling point is normally distributed, the distribution of the random work $\it f_{W {_{R}}}(w)$ is a chi-square function
\begin{eqnarray}
f_{W{_{R}}}(w)=(\sigma _x \cdot \sqrt{2 \pi \cdot k \cdot w})^{-1} \cdot exp(-(w/k)/2 \sigma^{2}_{x}),
\end{eqnarray}
with a standard deviation
\begin{eqnarray}
{\sigma} _{W{_{R}}}=\sqrt{3} \cdot k \cdot \sigma ^{2}_{x}=\sqrt{3} \cdot k \cdot m^{2} \cdot 2D \Delta t.
\end{eqnarray}

Jarzynski relation should be able to give free energy difference no matter what kind of work distribution guides a system between two states, but empirical results show that the work distribution properties (variance and distribution function) have strong influence on the convergence of the estimate when the number of samples is limited \cite{ps2004,grb2003}. If work variation (natural or externally induced) is greater than the difference between the average work and $\it \Delta G$, Jarzynski relation may underestimate free energy difference (Fig. 1, curves 4 \& 5) and \cite{pm2007}; if work variation is too small, it can not reduce the bias. Our analysis shows that for a modest number of samples (between 20 and 200 trajectories) the standard deviation $\it \sigma _{W{_{R}}}$ of the additional random work has to be close to the Jarzynski bias based on normal pulling ($\it\sigma _{W{_{R}}}\approx bias $) to be able to reduce it. In that case, Eq. 3 can be used to obtain the noise multiplication factor $\it m$ needed to attain such a work variation
\begin{eqnarray}
m\approx \sqrt{bias/(1.73\cdot k\cdot 2D\Delta t)}.
\end{eqnarray}
To obtain the maximum bias along the pulling coordinate we applied Eq. 9 from \cite{grb2003} which connects the maximum fluctuation(variation) of the estimate (Fig. 2a, curve 3) to its bias $\it\sigma^{2}_{J}=Var(e^{-\beta W_{dis}})/\beta ^{2}N=2\cdot bias(N)/\beta$.

For a typical SMD setup (k = 28 N/m, $\it D$ = 1.035$\cdot$$10^{-11}$$m^{2}$$s^{-1}$ \cite{mb1997}), time step $\Delta t$ = $10^{-13} s$ and maximum bias = $20 k_{B} T$, the above described procedure estimates the multiplication factor $\it m$ to be around 28 ($\it \sigma_x$ = 0.4\AA). For a bigger bias, $100 k_{B} T$ the same procedure estimates $\it m$ to be 64 ($\it \sigma_x$ = 0.87\AA). The effect of bias reduction is not very sensitive to the exact value of $\it m$, therefore we applied rounding of the calculated factor $\it m$ to the nearest lower decade.

The PMF estimates based on the constant velocity pulling (0.6 m/s) with external noise shown on Fig. 1 are averages of 10 Jarzynski reconstructions each based on 20 trajectories. The first estimate (Fig. 1, curve 4) was obtained with $\it m$ = 50 and the second with $\it m$ = 80 (Fig. 1, curve 5).

The additional noise helps in decreasing the overall bias of the PMF estimate between the initial and final state with a much smaller number of work trajectories but generates an underestimate when a random work deviation is greater than the bias. The additional noise with the smaller standard deviation ($\it m$ = 50) decreases the overall difference of the estimate and PMF (Fig. 1, curve 4) but overestimates $\it\Delta G$ between two equilibrium states. The noise with the greater deviation ($\it m$ = 80) decreases maximum bias much more efficiently but generates significant underestimate along the pulling trajectory (Fig. 1, curve 5). Those results show that a simple addition of the external noise can not consistently improve Jarzynski-based PMF calculation.

The PMF underestimate can be reduced if the external noise is adapted to the behavior of the bias along the reaction coordinate. The behavior of the bias is reflected in the estimate fluctuations along the reaction coordinate \cite{grb2003,ps2004}. The difference between the reconstruction and its smoothed version gives those fluctuations $\Delta \hat{G}_{noise} (r)$ (Fig. 2a). To get a smoothed version of the estimate we applied a simple low-pass filtering technique using a fifth-order Butterworth filter \cite{am2004} with the normalized cutoff frequency 10 for 500 samples along the reaction path and 20dB stop band attenuation; the rest of the harmonic spectrum has two orders of magnitude smaller amplitude and thus belongs to the fluctuations. To examine the behavior of the estimate's fluctuations we calculated 5 reconstructions per pulling velocity; each of them was based on 20 trajectories. The absolute, normalized, average version of the estimate fluctuations $\left|\Delta \hat{G}_{noise} (r)\right|/max(\left|\Delta \hat{G}_{noise}\right|)$ and their filtered variant $\it V_{noise}(r)$ are shown on Fig. 2b in comparison to the original, normalized PMF (pulling velocity 0.6 m/s). 

\begin{figure}
\includegraphics{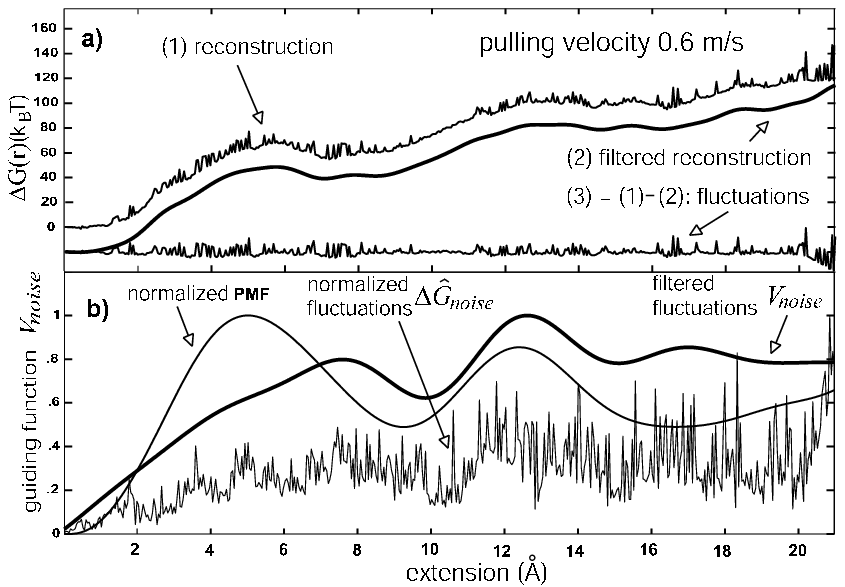}
\caption{\label{fig:figure_2}a) Reconstruction, filtered reconstruction and their difference. b) Average absolute normalized fluctuations of 5 reconstructions based on the pulling velocity 0.6 m/s compared to the normalized PMF. Thick line is $\it V_{noise}(r)$.}
\end{figure}

We developed two adaptive stochastic perturbation (ASP) protocols which use position dependent function $\it V_{noise}(r)$ to adjust the noise to the bias. The first protocol is $\it amplitude$ $\it modulation$ (AM) and it modulates the standard deviation of the applied noise. The second protocol modulates the frequency of the noise appearance so it is named $\it frequency$ $\it modulation$ (FM).

The amplitude modulation multiplies factor $\it m$ with the current value of the function $\it V{_{noise}}(r)$ as a way to improve the free energy reconstruction in a position dependent fashion. In this case the effective standard deviation of the additional noise $\it \sigma _{x}$ is not constant but depends upon the pulled coordinate $\it r(t)$, $\it \sigma _{x} =(\it V_{noise}(r)\cdot m)\cdot \sigma _{r} = (\it V_{noise}(r) \cdot m)\cdot \sqrt{2D\Delta t}$. Dotted curves on Fig. 3 show reconstructions based on two different pulling velocities, 0.2 m/s and 0.6 m/s. For both velocities we calculated $\it V_{noise}(r)$ and $\it m$ using reconstructions based on the normal pulling. The final PMF estimate was calculated as an ordinary average of 10 reconstructions (200 trajectories each). The thin profile on each subplot is reconstruction based on the normal pulling with the same number of trajectories (2000). Those results show that AM perturbation protocol can decrease bias without a significant underestimate.

The second method used to adapt noise is based on the modulation of the frequency of its application (FM). The noise in that case is not applied uniformly in time but its appearance depends on $\it V_{noise}(r)$ via the output of an additional random generator applied at the every time step of the simulation. This generator produces uniform random numbers between 0 and 1; the additional noise is applied only if the output of this generator is smaller than the current value of $\it V_{noise}(r)$. When applying this protocol we used the same values of the multiplication factor $\it m$ and the number of trajectories as in AM case. The dashed lines on Fig. 3 show the efficiency of the FM protocol in improving the Jarzynski estimate. It can be clearly seen that this protocol is able to reduce the bias with a minimal underestimate.
\begin{figure}
\includegraphics{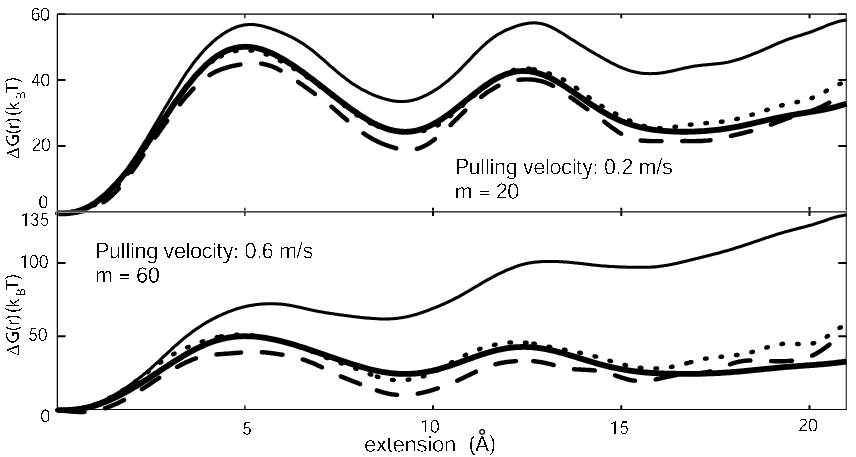}
\caption{\label{fig:figure_3}The PMF estimate based on the normal pulling (thin line) in comparison to the estimates based on amplitude (AM - dotted line) and frequency (FM - dashed line) modulated noise. Thick line is original PMF.}
\end{figure}

Fig. 4 shows the behavior of the estimate's RMSD (root mean square deviation, expressed as the percentage of the barrier height) for both modulation protocols and for normal pulling. When pulling velocity is 0.2 m/s and number of samples is limited (20$\sim$200), the bias is close to $20 k_{B}T$. The corresponding value of $\it m$ is 28, but we used three values of this factor (10, 20 and 30) to test the effects of low and high noise. For the faster pulling velocity (0.6 m/s) we also conducted experiments with three values of $\it m$ (40, 50 and 60) instead of using an exact value ($\it m$ = 67 for bias $\sim 110 k_{B}T$ based on 2000 trajectories). Fig. 4 shows that the additional noise can significantly decrease the bias along the reaction coordinate. It also depicts the undesired effects, an increase of RMSD coming from the underestimate when the random work is much larger than the average bias, i.e. when pulling is slow or number of samples is large enough to solely decrease the bias without the additional noise.

\begin{figure}[h]
\includegraphics{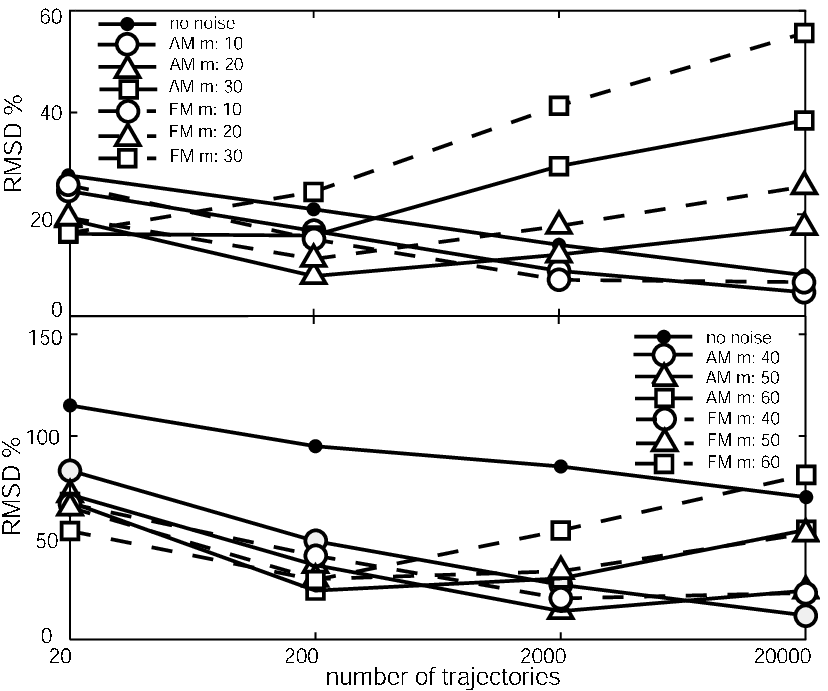}
\caption{\label{fig:figure_4}RMSD for three different values of the factor $\it m$ for both noise modulation protocols. Every point is an average of 10 reconstructions.}
\end{figure}

Both ASP protocols are able to decrease the number of samples needed to achieve the acceptable accuracy of the reconstruction expressed through RMSD between a given PMF and its estimate. For pulling velocity such as 0.2 m/s, the same RMSD (less than 10 \% of the barrier height) obtained using 20000 trajectories without an additional noise can be achieved with only 200 trajectories and noise modulation. For a faster pulling, 0.6 m/s, the improvement is even better because the quality of the reconstruction obtained with 200 or 2000 trajectories and noise modulation can not be achieved with 20000 normally pulled trajectories.

ASP is different from periodic loading \cite{bhs2004}, reversible pulling \cite{kbj2006} and improved sampling strategies based on random jumps from the trial trajectory \cite{sxs2003,yz2004}. Those approaches either require a memorizing of the current conformation which makes the whole procedure impossible with AFM experiments or they require going backward through the energy landscape, a difficult task to perform when proteins are pulled fast \cite{jk2004}. The last feature is very important in protein manipulation because proteins can not refold instantly, i.e. they need much more time to refold spontaneously than to unfold mechanically \cite{jk2004}. The noise adaptation protocols perform excellent in this aspect because they additionally decrease the probability of a sudden unfolding at the beginning of the perturbation process when polymer, i.e. protein is in the folded state. The skewed momenta protocol \cite{mfa2005} is similar to ASP but it directly introduces fluctuation to the pulled point and does not adjust it to the bias, and therefore does not avoid bias. Both noise modulation protocols can be modified to PMF calculation in other types of physical and chemical experiments with suitable modulation of the corresponding reaction coordinates. Our current research is focused on the more efficient estimation of the noise guiding function $\it V_{noise}(r)$.

\bibliography{jarz_arXiv_sf}

\end{document}